\documentclass[twocolumn,aps]{revtex4}

\usepackage{graphicx} 

\begin{document}

\title{2-D color code quantum computation}

\author{Austin G. Fowler}
\affiliation{Centre for Quantum Computer Technology, The University
of Melbourne, Victoria 3010, Australia}

\date{\today}

\begin{abstract}
We describe in detail how to perform universal fault-tolerant
quantum computation on a 2-D color code, making use of only nearest
neighbor interactions. Three defects (holes) in the code are used to
represent logical qubits. Triple defect logical qubits are deformed
into isolated triangular sections of color code to enable
transversal implementation of all single logical qubit Clifford
group gates. CNOT is implemented between pairs of triple defect
logical qubits via braiding.
\end{abstract}

\maketitle

\section{Introduction}

Classical computers manipulate bits that can be exclusively 0 or 1.
Quantum computers manipulate quantum bits (qubits) that can be
placed in arbitrary superpositions $\alpha|0\rangle +
\beta|1\rangle$ and entangled with one another to create states such
as $(|00\rangle + |11\rangle)/\sqrt{2}$. This additional flexibility
provides both additional computing power
\cite{Shor94b,Grov96,Subr02,Cheu09,Harr09,Lloy96,Wieb10,Smit10} and
additional challenges when attempting to cope with the now quantum
errors in the computer \cite{Shor95,Cald95,Stea96}. An extremely
efficient scheme for quantum error correction and fault-tolerant
quantum computation is required to correct these errors without
making unphysical demands on the underlying hardware and without
introducing excessive time overhead and thus wasting a significant
amount of the potential performance increase.

Recently, significant progress towards practical quantum error
correction and fault-tolerant quantum computation has been made by
making use of topological error correction \cite{Raus07,Raus07d,
Bomb07,Fowl08,Wang10}.  These schemes feature a single error
correcting code used for the entire computer with qubits associated
with holes or ``defects'' deliberately introduced using
measurements.  Logical qubits can be initialized and measured in the
$X_L$ and $Z_L$ bases.  Logical CNOT involves braiding defects
around one another.  Individual logical qubits can be isolated and
some transversal single logical qubit operations applied
\cite{Bomb07,Fowl08}.  All of these schemes possess a sufficiently
broad range of gates to enable state distillation
\cite{Brav05,Reic05} and thus achieve universality, but none possess
a sufficiently broad range of gates to enable universal computation
without state distillation. Indeed, it has recently be suggested
that it is not possible for these types of topological schemes to
avoid non-topological techniques, such as state distillation, to
enable universal quantum computation \cite{Sarv10}.

State distillation is typically used to produce a better copy of one
or both of the states $|Y\rangle=(|0\rangle+i|1\rangle)/\sqrt{2}$
and $|A\rangle=(|0\rangle+e^{i\pi/4}|1\rangle)/\sqrt{2}$ consuming
either 7 or 15 imperfect copies of these states, respectively. Given
reasonable assumptions about the desired logical error rate and the
underlying physical error rate, three or more concatenated layers of
state distillation can easily be required to produce sufficiently
high fidelity states \cite{Raus07d}.  A single Toffoli gate requires
7 accurate copies of $|A\rangle$ and up to 7 accurate copies of
$|Y\rangle$ \cite{Raus07d,Niel00}.  Depending on the
\begin{figure}
\begin{center}
\resizebox{75mm}{!}{\includegraphics{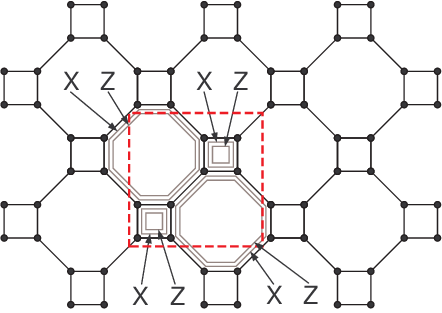}}
\end{center}
\caption{The stabilizers of the error correction substrate.  Dots
indicate the location of qubits.  The dashed box is a unit cell of
the lattice containing 8 qubits and associated with 8 independent
stabilizers.} \label{stabilizers}
\end{figure}
details of the quantum algorithm being executed, the ancilla factory
required to produce a sufficiently high rate of distilled states can
easily be several orders of magnitude larger than the rest of the
computer. Reducing the reliance on state distillation can thus
result in a large reduction of the required number of qubits.

In this work, we combine a 2-D color code \cite{Bomb06} with defect
braiding, defect isolation and transversal rotation to enable the
implementation of CNOT and the entire single qubit Clifford group of
gates. This problem has also received attention in a recent work
\cite{Bomb10}, however the scheme presented here is simpler. Our
scheme calls for a 2-D array of qubits with local tunable
interactions and a measurement time of the same order as the gate
times. It features a reasonably high threshold error rate of
approximately 0.1\% \cite{Wang09b}. Our scheme also supports fast
long-range logical gates and relatively low qubit overhead due to
both its use of efficient topological error correction and its
reduced reliance on state distillation.

The discussion is organized as follows.  In Section~\ref{Error
correction substrate} we review a 2-D color code from \cite{Bomb06}
which forms the error correction substrate of all that follows.
Logical qubit initialization and measurement are described in
Section~\ref{Logical qubit initialization and measurement}, with
each logical qubit being represented by three defects.
Section~\ref{Defect deformation} describes defect deformation.  The
\begin{figure*}
\begin{center}
\resizebox{120mm}{!}{\includegraphics{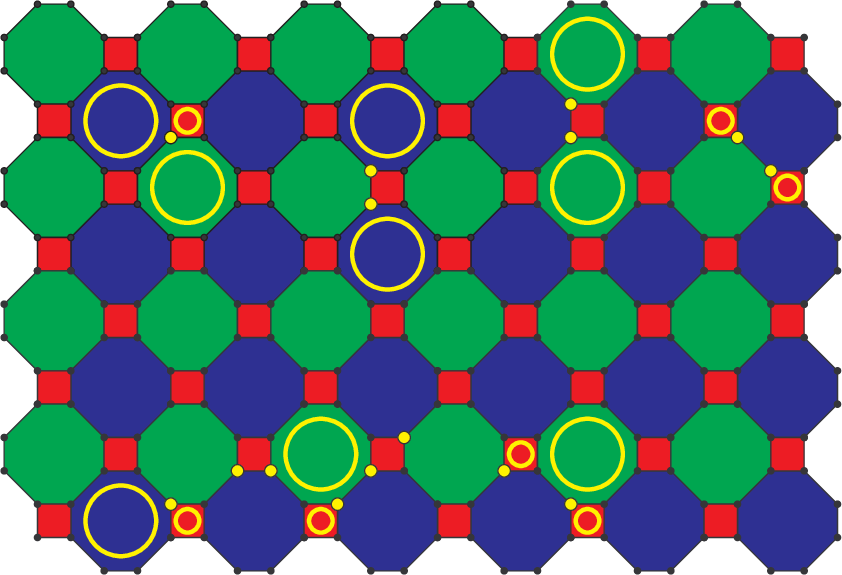}}
\end{center}
\caption{Examples of errors in a color code (color online). Yellow
dots indicate the locations of errors.  Yellow circles indicate
stabilizers of changed sign. As drawn, each chain of errors is of a
single type $X$ or $Z$ and the neighboring stabilizers of changed
sign are of type $Z$ or $X$, respectively.} \label{errors}
\end{figure*}
logical gates CNOT, $H$, $X$, $Z$ and $S$ are detailed in
Section~\ref{Basic logical gates}. Section~\ref{Conclusion}
summarizes our results.

\section{Error correction substrate}
\label{Error correction substrate}

Consider Fig.~\ref{stabilizers}.  This shows a 2-D lattice of qubits
arranged on faces with either 4 or 8 edges \cite{Bomb06}.  Each face
is associated with two stabilizers \cite{Gott97}: the tensor product
of $X$ on every qubit around the face and similarly for $Z$.  Note
that because every face has two qubits in common with its
neighboring faces, all stabilizers commute.  Note also that the unit
cell indicated in Fig.~\ref{stabilizers} contains 8 qubits and can
be associated with 8 independent stabilizers.  An infinite lattice
of this form therefore contains no logical qubits.  In the absence
of errors, the lattice of qubits may be assumed to be in the
simultaneous $+1$ eigenstate of each stabilizer.

Fig.~\ref{errors} contains examples of the effects of errors.  Note
that one of the three colors red, blue and green has been assigned
to each of the faces such that no two adjacent faces have the same
color.  This will simplify the discussion of the various types of
errors and the later discussion of logical operators.  Every qubit
is on three faces.  If a qubit suffers an $X$/$Z$ error, the $Z$/$X$
stabilizers of these three faces become negative if no other error
of the same type occurs on these three faces.  A second error of the
same type adjacent to the first error results in just two faces of
the same color having negative stabilizers.  Such chains of errors
are said to have the same color as the faces they connect.  A chain
of each color can meet at a single qubit without changing the sign
of any stabilizers.  Arbitrarily complex error trees can occur with
multiple intersections.

Additional syndrome qubits are required to determine
\begin{figure}
\begin{center}
\resizebox{63.5mm}{!}{\includegraphics{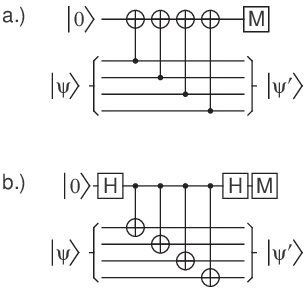}}
\end{center}
\caption{Circuit showing how an additional syndrome qubit (top line
of each figure) is used to measure red face a.) $Z$ stabilizers, b.)
$X$ stabilizers.} \label{syndrome_circuits}
\end{figure}
the sign of the face stabilizers.  For red faces we choose to use
just one additional syndrome qubit and the simple circuits shown in
Fig.~\ref{syndrome_circuits} to determine the sign of their
associated $X$ and $Z$ stabilizers.  Note that errors occurring
during these circuits can propagate to the data qubits with a single
syndrome qubit error propagating to multiple data qubits.  For red
faces, the potential number of effected data qubits is sufficiently
low that we choose to leave the detection and correction of these
errors to later rounds of syndrome extraction.

\begin{figure}
\begin{center}
\resizebox{60mm}{!}{\includegraphics{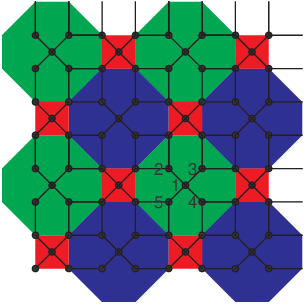}}
\end{center}
\caption{Underlying lattice of physical qubits.  Dots represent
qubits, lines represent tunable interactions between qubits.  The
numbered qubits are used in
Figs.~\ref{x_syndrome}--\ref{z_syndrome}.} \label{syndrome_qubits}
\end{figure}

\begin{figure}
\begin{center}
\resizebox{70mm}{!}{\includegraphics{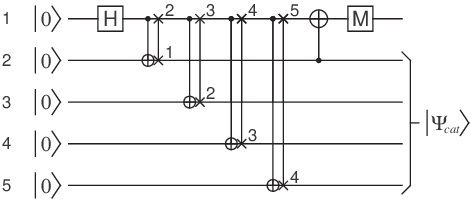}}
\end{center}
\caption{Preparation of a 4-qubit cat state using the 5 qubits
indicated in Fig.~\ref{syndrome_qubits}.  The states initially
stored in qubits 1--4 are manipulated to form a cat state stored on
qubits 2--5.  The state stored in qubit 5 is used to check the cat
state and is read out using qubit 1.} \label{cat}
\end{figure}

\begin{figure}
\begin{center}
\resizebox{85mm}{!}{\includegraphics{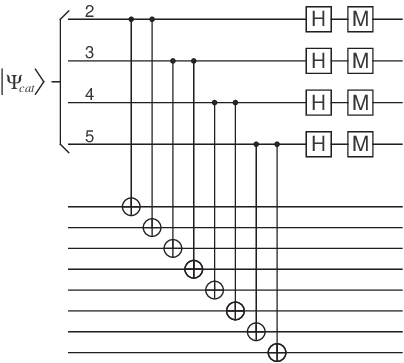}}
\end{center}
\caption{Circuit used to determine the eigenvalue of the 8 qubit $X$
stabilizer associated with a green or blue face.} \label{x_syndrome}
\end{figure}

\begin{figure}
\begin{center}
\resizebox{85mm}{!}{\includegraphics{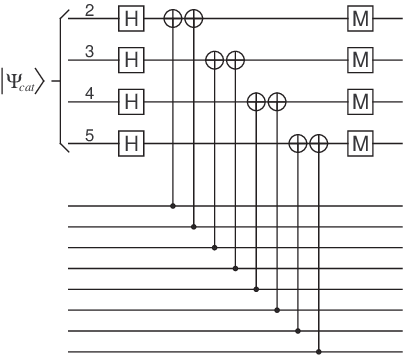}}
\end{center}
\caption{Circuit used to determine the eigenvalue of the 8 qubit $Z$
stabilizer associated with a green or blue face.} \label{z_syndrome}
\end{figure}

\begin{figure}
\begin{center}
\resizebox{65mm}{!}{\includegraphics{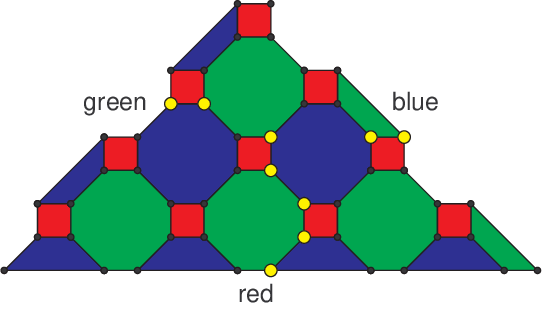}}
\end{center}
\caption{Examples of the three colors of boundaries and error chains
of each color starting at each boundary and meeting a single qubit.}
\label{borders}
\end{figure}

Green and blue faces, with 8 data qubits, could simply use the 8
qubit analogues of Fig.~\ref{syndrome_circuits} and live with the
fact that up to 4 data qubits could be effected by an error on the
single syndrome qubit.  We choose not to do this.  Instead, 5
syndrome qubits are devoted to each green and blue face as shown in
Fig.~\ref{syndrome_qubits}.  Initially, the circuit of
Fig.~\ref{cat} is repeatedly executed until the central syndrome
qubit is measured in state $|0\rangle$ indicating successful
preparation of a 4-qubit cat state provided no more than one error
occurred during the execution of the circuit.  The circuits of
Fig.~\ref{x_syndrome} and Fig.~\ref{z_syndrome} are then executed
for $X$ and $Z$ syndrome extraction respectively with each qubit of
the cat state interacting with its two non-syndrome nearest
neighbors.

In addition to reducing the number of data qubits that can be
corrupted after a single error during syndrome extraction, a
significant benefit of using 5 syndrome qubits on green and blue
faces is avoiding the need to have a single syndrome qubit coupled
to 8 data qubits, greatly
\begin{figure*}
\begin{center}
\resizebox{120mm}{!}{\includegraphics{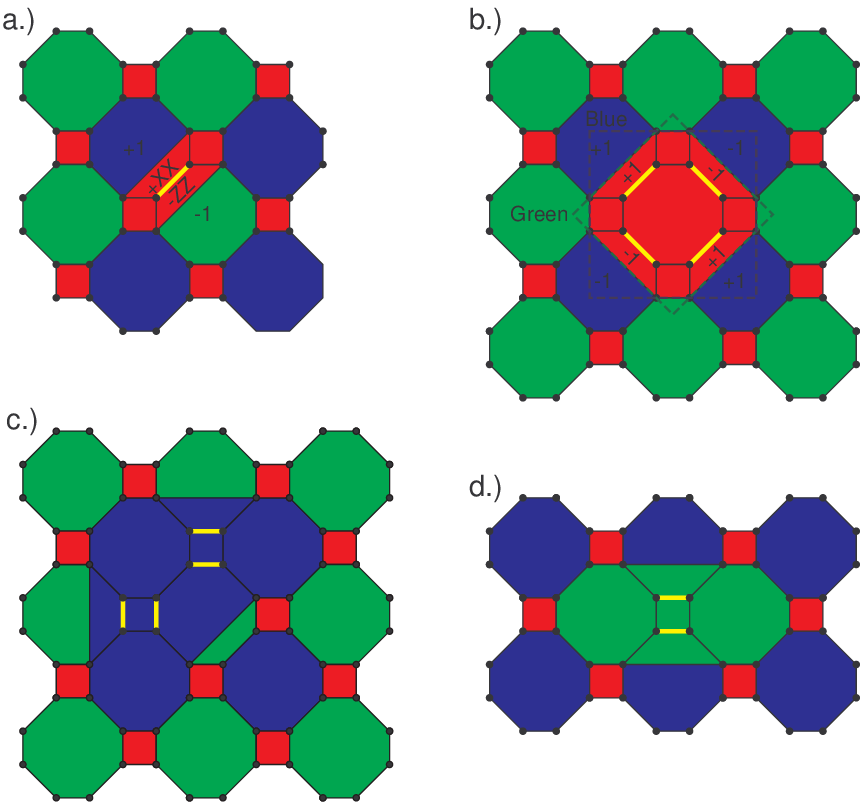}}
\end{center}
\caption{a.) Red defect created by measuring $XX$ and $ZZ$ on the
qubits indicated by a yellow line (color online).  The signs of
these measurements determine the signs of the stabilizers of the
neighboring faces of reduced size.  b.) Red defect created by
measuring a complete face.  Equivalent blue and green boundary
stabilizers with positive sign independent of the measurement
results assuming no errors. These boundary stabilizers are always
positive as they are equal to the product of the five complete face
stabilizers contained within the effect. c.) Blue defect. d.) Green
defect.} \label{defects}
\end{figure*}
\begin{figure*}
\begin{center}
\resizebox{150mm}{!}{\includegraphics{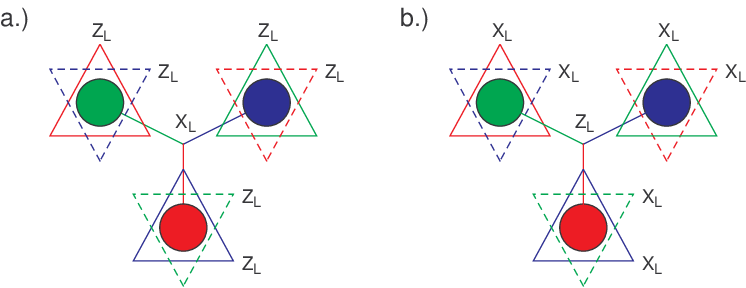}}
\end{center}
\caption{A defect of each color forming a.) a primal qubit, b.) a
dual qubit.  Note that the six primal $Z_L$ and dual $X_L$ operators
are equivalent.} \label{logical qubits}
\end{figure*}
simplifying the underlying lattice of
qubits and network of connections.  Note that in
Fig.~\ref{syndrome_qubits} no qubit is connected to more than 4
other qubits with most connected to just 3 other qubits.  Note also
that the concepts of color and the different face sizes arise only
from how the underlying hardware is used, not from any aspect of its
physical construction.

The lattices we will discuss in this work will both have and make
extensive use of boundaries.  Fig.~\ref{borders} shows examples of
the three different colors of boundaries.  At this point in the
discussion, the only property of a boundary of a given color that is
of interest is the fact that an error chain of the same color can
connect to it without changing the sign of any stabilizers.  An
example of an error chain of each color starting at each boundary
and meeting at a single qubit is shown. The details of how error
correction might be performed and a calculation of the threshold
error rate can be found in \cite{Wang09b}.

\section{Logical qubit initialization and measurement}
\label{Logical qubit initialization and measurement}

In Section~\ref{Error correction substrate} we discussed an infinite
lattice of qubits and error correction circuits, however the lattice
contained no logical qubits.  Logical qubits can be introduced by
ceasing to enforce stabilizers and thereby introducing degrees of
freedom into the lattice.  We call a connected region of a single
color of faces whose stabilizers we no longer enforce a defect.
Examples of red, green and blue defects are shown in
Fig.~\ref{defects}.  Note that by connected we mean faces connected
by an $XX$ and a $ZZ$ measurement.  The effect of these measurements
is to create a single large face of the same color as the
constituent faces.  Initially, provided the defect is constructed
from the measurement of complete faces, it is in the $+1$ eigenstate
of its associated bounding $X$ and $Z$ stabilizers. We will
henceforth always use defects constructed from the measurement of
complete faces.  Note that the bounding stabilizers can be regarded
as rings of either of the two colors the defect isn't.

Before a defect can be used as part of a logical qubit, some of its
neighboring stabilizers must be corrected.  Stabilizers that have
had their sides reduced by $XX$ and $ZZ$ measurements will have a
sign dependent on the results of these measurements.  Note that such
reduced size stabilizers of negative sign always occur in pairs.
They can therefore also be connected and corrected in pairs using
appropriate chains of operators.

Note that direct $XX$ and $ZZ$ measurements are actually not
necessary to create a large defect.  It is equivalent and simpler to
just measure the stabilizers of reduced size.  The qubits inside the
defect can be ignored.  These inner qubits only become important
again if the size of the defect is reduced.

Many types of logical qubits are possible.  We will be interested in
logical qubits consisting of three defects, one of each color. Note
that chains of operators of the same
\begin{figure}
\begin{center}
\resizebox{80mm}{!}{\includegraphics{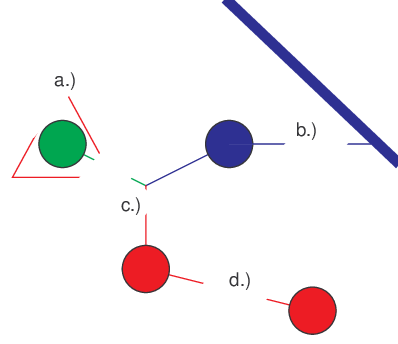}}
\end{center}
\caption{Examples of error chains likely to lead to logical errors.
a.) Half encircling a defect.  b.) Half connecting a defect to a
boundary of the same color.  c.) Half connecting the three defects.
d.) Half connecting a defect to another of the same color.}
\label{logical errors}
\end{figure}
color commute regardless of
whether they consist of $X$ or $Z$ operators as they always have an
even number of qubits in common.  Fig.~\ref{logical qubits} shows
examples of the two types of triple defect logical qubits we will
use.  We call logical qubits with triangular $X_L$ operators
``primal'' and those with triangular $Z_L$ operators ``dual''.
Primal qubits will store the data in our computer whereas dual
qubits will facilitate multiple qubit gates. Note that the six types
of primal $Z_L$ operators are equivalent in the sense that they all
commute with one another and all anticommute with primal $X_L$. When
performing a logical phase-flip, it does not matter which primal
$Z_L$ operator is used.  Dual $X_L$ operators are also all
equivalent.

Primal qubits are naturally initialized to the $+1$ eigenstate of
$Z_L$, namely $|0_L\rangle$.  Similarly, dual qubits are naturally
initialized to $|+_L\rangle$.  Note that since primal and dual
qubits are structurally identical, when we create a logical qubit we
actually create both a primal and a dual qubit.  Furthermore, we do
nothing to distinguish these two types of qubits.  As we shall see,
computation proceeds with both types of qubit present and we simply
ignore one type and focus on the computation occurring in the qubit
type of interest.

Measurement of a logical qubit can be performed transversely by
measuring a region of qubits encompassing the three defects in
either the $X$ or $Z$ basis.  Measurement results are error
corrected using the standard algorithm with syndromes calculated
from the parity of measurement results around each face.  After
error correction, provided no logical errors have occurred, all
rings of measurements around each defect will have the same parity
and similarly all three-way chains of measurements connecting all
three defects will have the same parity.
\begin{figure*}
\begin{center}
\resizebox{150mm}{!}{\includegraphics{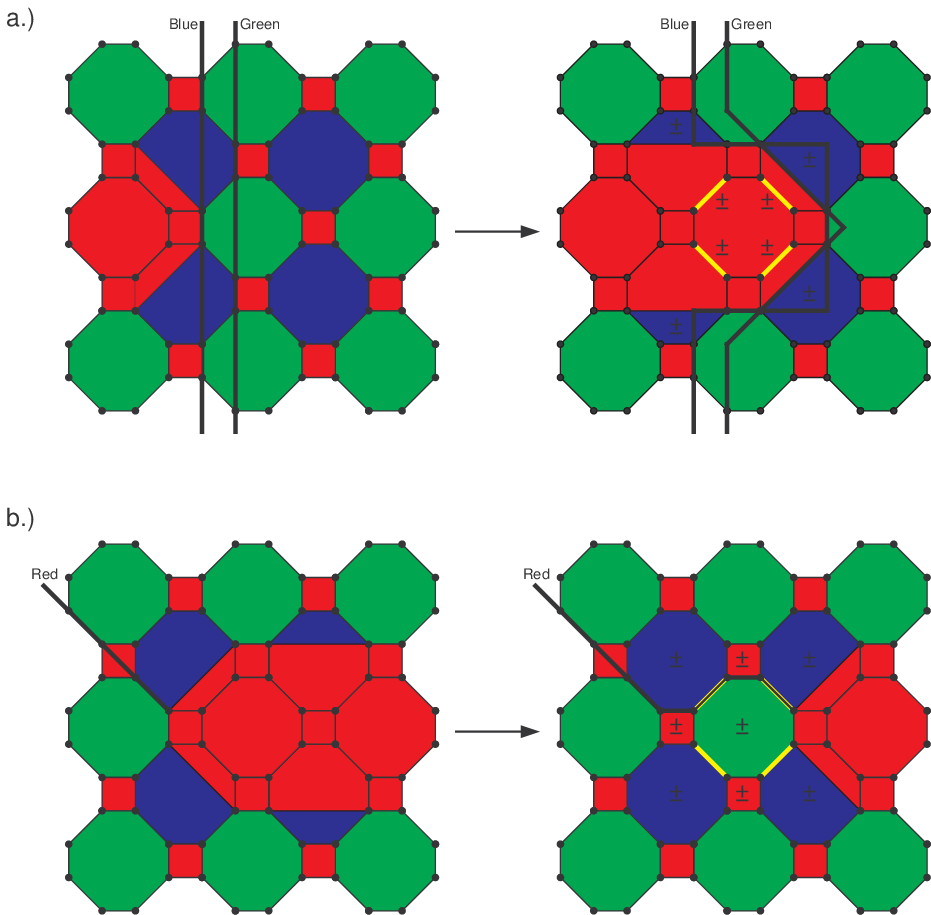}}
\end{center}
\caption{a.)  Procedure for expanding a red defect. It is
conceptually simpler to imagine that the operators $XX$ and $ZZ$ are
measured directly on the qubits on the yellow lines (color online),
but in practice these qubits can be ignored and only the stabilizers
of the indicated partial faces measured. b.) Procedure for
contracting a red defect.  The full stabilizers of the indicated
faces are measured once more. The signs of these face stabilizers
are then corrected using the regular error correction procedure.}
\label{defect_movement}
\end{figure*}
These parities are the
logical measurement results.  Note that when performing the
transversal $X$/$Z$ measurements we perform both a primal and a dual
$X_L$/$Z_L$ measurement.  The unnecessary logical measurement result
is simply ignored.

Primal/dual qubits can also be initialized to
$|+_L\rangle$/$|0_L\rangle$ by first initializing a region of qubits
to $|+\rangle$/$|0\rangle$ and then creating defects by measuring
appropriate $Z$/$X$ stabilizers.  These stabilizers will have random
sign initially, and must be corrected before the logical qubit is
used.

A number of different types of logical errors are possible.
Fig.~\ref{logical errors} contains a few examples.  If an error
chain half encircles a defect, it cannot be reliably corrected as
given only endpoint information is not possible to know which half
of the defect is encircled and thus which half of the defect to
apply corrective operations to.  If the wrong half is chosen, we
form a logical operation instead of correcting the error.  For
primal/dual qubits, half rings of $Z$/$X$ errors are dangerous.
Similarly, if a collection of different color error chains half
constructs a three-way connection, the error correction procedure
cannot in general determine which half has been constructed and
correct it. For primal/dual qubits, half connections of $X$/$Z$
errors are dangerous.  Furthermore, individual defects of a given
color can be half connected to other defects of the same color by
error chains or half connected to boundaries of
\begin{figure*}
\begin{center}
\resizebox{120mm}{!}{\includegraphics{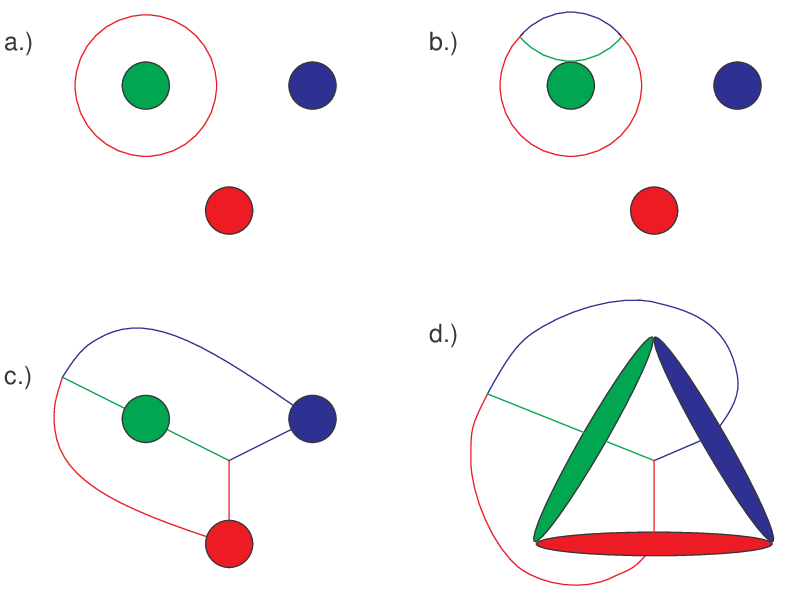}}
\end{center}
\caption{Step-by-step deformation of a ring operator into a pair of
tree operators via deformation of the associated defects to create
an isolated region of lattice.} \label{equivalence}
\end{figure*}
\begin{figure*}
\begin{center}
\resizebox{170mm}{!}{\includegraphics{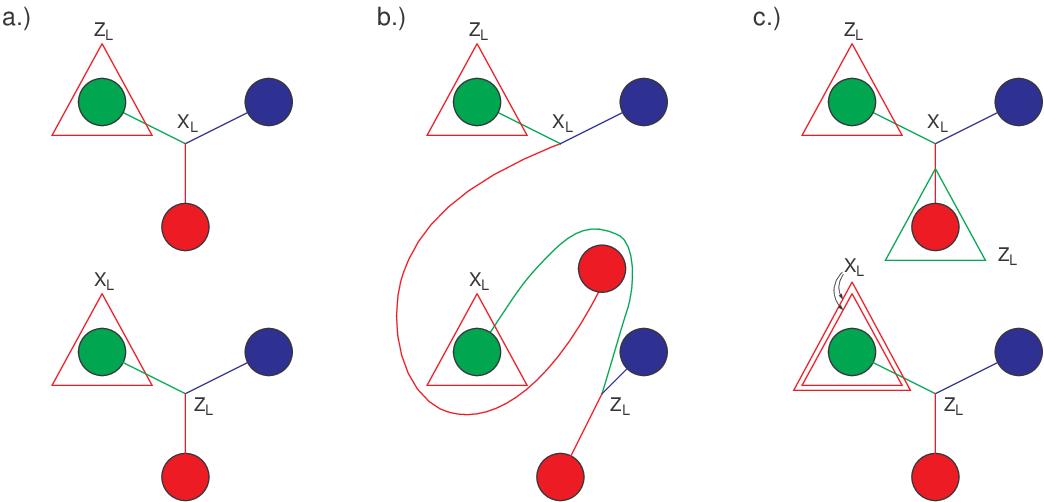}}
\end{center}
\caption{Logical CNOT via braiding with a primal qubit (top) as
control and dual qubit (bottom) as target. The appropriate mappings
of logical stabilizers $XI\mapsto XX$, $IX\mapsto IX$, $ZI\mapsto
ZI$ and $IZ\mapsto ZZ$ can be seen by tracing the deformation of one
logical stabilizer at a time.} \label{CNOT}
\end{figure*}
the same color. Both situations correspond to errors that cannot be
reliably corrected. Given the rich error structure of color codes,
an unlimited number of additional types of logical errors are
possible, though suppressed if boundaries and defects are kept well
separated. Note that the strength of $X$/$Z$ error correction can be
independently set by adjusting the circumference and separation of
defects.

\section{Defect deformation}
\label{Defect deformation}

We have discussed error correction and the creation and measurement
of logical qubits.  We now turn our attention to the techniques
required to perform computation.  Defect expansion and contraction
can be used to adjust the size of a defect and thus the local error
correction strength.  By combining expansion and contraction,
defects can be moved and braided around one another, realizing
multiple qubit gates as we shall see in Section~\ref{Basic logical
gates}.  Expanding all three defects comprising a logical qubit
until they touch and creating an enclosed triangular region isolates
the logical qubit from the rest of the lattice, enabling transversal
gates to be applied.

Consider Fig.~\ref{defect_movement}a.  This shows the procedure for
expanding a red defect and the effect of doing so on both a green
and blue stabilizer.  Fig.~\ref{defect_movement}b shows the effect
of contracting a defect on a red stabilizer attached to the defect.
Note that even in the absence of errors, corrective operations must
be applied when moving the defect to ensure that the signs of all
stabilizers remain unchanged at the end of the procedure.  The
procedures for expanding and contracting green and blue defects are
analogous.

Fig.~\ref{equivalence} shows the effect of isolating a triangular
region of the lattice, the detailed structure of which can be found
in Fig.~\ref{borders}, by simultaneously expanding all three defects
comprising a logical qubit.  Stabilizers encircling a defect are
converted into pairs of three-way stabilizers. Section~\ref{Basic
logical gates} makes use of such isolated regions.  Note that the
isolation is reversible by simply contracting the defects once more.

\section{Logical gates}
\label{Basic logical gates}

In this section, we describe the logical gates CNOT, $H$, $X$, $Z$
and $S$.  By far the simplest logical gates are $X_L$ and $Z_L$,
which can be implemented with simple rings and three-way trees of
single-qubit operators.  We shall not discuss these further.  The
remaining logical gates require individual discussion.

Logical Hadamard, $H_L$, shall be applied transversely to an
isolated triangular logical qubit, however some care is required.
Given a primal/dual qubit, during the isolation process $Z_L$/$X_L$
is split into two three-way trees.  The tree external to the
isolated region must be removed before transversal gates can be
applied.  This can be achieved by measuring a region of qubits
around the isolated region in the $Z$/$X$ basis.  Note that all
external operator trees must have the same parity in the absence of
errors and that the standard error correction procedure can be used
to ensure that the parity is determined fault-tolerantly.  Note that
this does not constitute logical measurement of $Z_L$/$X_L$ as the
isolated region is not measured and the parity of internal three-way
trees must also be known to effect logical measurement. Measuring
the parity $s$ of external trees does, however, introduce a
byproduct operator $X^s_L$/$Z^s_L$. With the external trees pruned,
logical Hadamard can then be applied transversely.  The appropriate
external face stabilizers must be measured and corrected before the
logical qubit is converted back to three isolated defects.  The
conversion back can also introduce a byproduct operator.

Logical $S$ requires a different type of care.  In this case, the
external trees do not need to be pruned, however we need to ensure
that the correct number of qubits have been enclosed.  In the
original color code work, triangular logical qubits with $3 {\rm
~mod~} 4$ total physical qubits and either 4 or 8 qubits per face
were used \cite{Bomb06}.  For convenience, we require there to be $1
{\rm ~mod~} 4$ enclosed qubits.  Using enclosed regions of the form
shown in Fig.~\ref{borders}, this can be achieved by having an even
number of rows of red faces.  When single-qubit $S$ is applied
transversely, the condition of having either 4 or 8 qubits per face
ensures that every state comprising $|0_L\rangle$ and every state
comprising $|1_L\rangle$ acquires the same phase. Having $1 {\rm
~mod~} 4$ qubits in total ensures that
$|0_L\rangle\mapsto|0_L\rangle$ and $|1_L\rangle\mapsto
i|1_L\rangle$. Transversal $S$ thus implements $S_L$.

With an odd number of enclosed qubits and every three-way chain
containing an odd number of operators, we can also implement $X_L$
and $Z_L$ transversely.  While this would not be done under normal
circumstances, as it is easier to simply apply ring or tree
operators to non-isolated logical qubits, if the logical qubit has
already been isolated, the ability to apply transversal $X_L$ and
$Z_L$ enables us to avoid further modification of the shape of the
logical qubit.

\begin{figure}
\begin{center}
\resizebox{85mm}{!}{\includegraphics{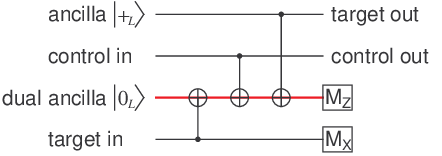}}
\end{center}
\caption{Circuit constructed from available components implementing
primal-primal logical qubit CNOT.} \label{CNOT_circuit}
\end{figure}

Logical CNOT is carried out in a similar manner to the schemes of
\cite{Raus07d,Fowl08}. A primal qubit can be used to control a CNOT
gate with a dual qubit as target as shown in Fig.~\ref{CNOT}.  Note
that all of the CNOT stabilizer mappings $XI\mapsto XX$, $IX\mapsto
IX$, $ZI\mapsto ZI$ and $IZ\mapsto ZZ$ are faithfully realized.  We
can then use the circuit shown in Fig.~\ref{CNOT_circuit} to
simulate CNOT between two primal qubits.

\section{Conclusion}
\label{Conclusion}

We have described how to perform universal fault-tolerant quantum
computation with reduced reliance on state distillation on a
specific 2-D color code. This scheme has a relatively high threshold
error rate of 0.1\%, relatively low qubit overhead, fast long-range
logical gates and makes few demands on the underlying hardware.  Our
scheme calls for a measurement time of the same order as the gate
times, local single qubit unitaries and a 2-D nearest neighbor
tunably coupled lattice of qubits, with each qubit coupled to either
3 or 4 neighbors and with no couplings crossing.  This lattice is
shown in Fig.~\ref{syndrome_qubits}.  Ideally, it should be possible
to simultaneously measure an arbitrary subset of qubits in the
lattice, although if measurement hardware cannot be located near
every qubit, it would be sufficient to be able to measure a fraction
of the qubits that is independent of the lattice size.

\bibliography{../References} 

\end{document}